\documentclass[11pt]{article}

\usepackage[a4paper,margin=3cm]{geometry}

\usepackage[parfill]{parskip}
\usepackage{setspace}
\usepackage[labelfont=bf,font={small,stretch=0.95}]{caption}
\usepackage[square,sort,compress,numbers]{natbib}
\usepackage{hyperref}

\usepackage{amsmath}
\usepackage{amssymb}

\usepackage{graphicx}
\usepackage{mathptmx}

\usepackage{multirow}

\usepackage{xcolor}
\usepackage{soul}

\begin{document}

\title{{\huge Synthetic pre-training for neural-network interatomic potentials\par}}

\author{{\Large John L. A. Gardner, Kathryn T. Baker, and Volker L. Deringer$^\ast$}\\[4mm]
        {\em Department of Chemistry, Inorganic Chemistry Laboratory,  }\\
        {\em University of Oxford, Oxford OX1 3QR, UK}\\[4mm]
        $^\ast$\href{mailto:volker.deringer@chem.ox.ac.uk}{volker.deringer@chem.ox.ac.uk}}

\maketitle

\begin{abstract}
\noindent
Machine learning (ML) based interatomic potentials have transformed the field of atomistic materials modelling. 
However, ML potentials depend critically on the quality and quantity of quantum-mechanical reference data with which they are trained, and therefore developing datasets and training pipelines is becoming an increasingly central challenge.
Leveraging the idea of ``synthetic'' (artificial) data that is common in other areas of ML research, we here show that synthetic atomistic data, themselves obtained at scale with an existing ML potential, constitute a useful pre-training task for neural-network interatomic potential models. 
Once pre-trained with a large synthetic dataset, these models can be fine-tuned on a much smaller, quantum-mechanical one, improving numerical accuracy and stability in computational practice.
We demonstrate feasibility for a series of equivariant graph-neural-network potentials for carbon, and we carry out initial experiments to test the limits of the approach.
\end{abstract}

\section{Introduction}
\label{sec:intro}

Machine-learning interatomic potential (MLIP) models are increasingly used to accelerate the simulation, discovery, and design of molecules and materials \cite{Behler-17-05, Deringer-19-09, Noe-20-02, Unke-21-03, Friederich-21-05}.
MLIPs approximate quantum-mechanical potential energies and forces acting on atoms, yet require orders of magnitude lower computational cost than the corresponding reference methods.
As such, they have begun to enable real-world applications that would otherwise have been out of reach for quantum-mechanically accurate simulations: the behaviour of matter under extreme conditions \cite{Cheng-20-09}; the complex atomic structure of amorphous solids \cite{Zhou-22-02}; the discovery of unconventional reaction mechanisms \cite{Westermayr-22-08}.

To further increase their impact, and to enable more widespread adoption and application in the natural sciences, it is necessary that non-specialists are able to quickly, cheaply, and yet reliably train MLIPs for new systems.
A major bottleneck in this endeavour is the computation of expensive quantum-mechanical reference data used in training.
Significant research efforts are therefore being spent on designing data-efficient MLIP fitting frameworks, and also on architecture-agnostic strategies that reduce the amount of training data required to reach a given level of accuracy.

In the fields of computer vision and natural language processing in particular, one strategy for creating specialised ML models is to leverage existing ``foundation'' models that have been pre-trained on a general task over very large amounts of data.
End users can then cheaply fine-tune such models for their desired domain and specific task using small amounts of data.
However, those approaches rely on the availability of large corpora of existing data for pre-training -- this is not often the case in atomistic ML if one is starting from quantum mechanics directly.

In the present work, we show that the prediction of synthetic energy and force labels, cheaply generated by existing MLIPs, is a useful pre-training task for neural-network (NN) interatomic potentials, enabling subsequent fine-tuning on quantum-mechanical data. Specifically:
\begin{itemize}
    \item We show proof-of-concept that synthetic pre-training of NN potentials for carbon improves model accuracy compared to direct training, particularly in the low-data regime;
    \item We show that the fine-tuned model accuracies are sensitive to the source of the synthetic data labels (that is, different pre-training methods are differently well suited for this task);
    \item We show that starting from general, synthetically pre-trained models and fine-tuning them for specific use cases improves the physical robustness of the final potentials, as measured by the ability to run stable dynamics outside the scope of the fine-tuning dataset;
    \item We suggest that MLIPs can be repurposed as alchemical pre-training sources: we show that pre-training models on synthetic data for carbon, with proportionally scaled structures, leads to positive transfer when fine-tuning on data for silicon.
\end{itemize}

\section{Related work}
\label{sec:related}

\textbf{Synthetic data.} The term ``synthetic'' refers to data that have been created with a surrogate model, for instance from simulations or generative modelling; this is in contrast to ``real'' data, which are obtained from physical measurements, computed using quantum mechanics (\emph{ab initio}), or collected from other reliable observations of the ground truth.
Increasingly, synthetic data are being used to (pre-) train ML models, most notably in the fields of image classification, segmentation and generation \cite{Azizi-23-04,Kirillov-23-04}, natural language modelling \cite{To-23-05}, and speech detection \cite{Zhang-22-07}. 

Synthetic data are also beginning to be used in the physical sciences:
\citeauthor{Aty-22-04} studied synthetic data as a pre-training task for determining experimental lipid phase behaviour from small-angle X-ray scattering patterns \cite{Aty-22-04};
\citeauthor{Anker-23-03} explored the use of synthetic data in interpreting inelastic neutron-scattering data \cite{Anker-23-03};
\citeauthor{Schuetzke-23-06} used synthetic data to mimic the characteristic appearance of experimental measurements for several spectroscopy methods \cite{Schuetzke-23-06}.

In the context of atomistic ML, we have recently shown that synthetic data allow one to distill reliable, but comparably expensive MLIPs into faster and cheaper ones in a teacher--student manner \cite{Morrow-22-09}, and that synthetic data provide a cheap means to explore atomistic energy models \cite{Gardner-22-12,FaureBeaulieu-23-05}. We showed initial proof-of-concept that synthetic data constitute a useful pre-training task, limited at that time to simple feed-forward NNs regressing scalar atomic quantities \cite{Gardner-22-12}. Very recently, \citeauthor{Kelvinius-23-06} investigated the use of synthetic data as a means to aid knowledge distillation from slow to fast graph-neural-network potentials via intermediate learned representations \cite{Kelvinius-23-06}.

\textbf{Pre-training tasks for MLIPs.}
Fine-tuning existing pre-trained NNs can improve accuracy and data efficiency compared to directly training models. 
As such, much research effort is expended on exploring useful, effective, and architecture-agnostic pre-training tasks. 
In the field of MLIP development, \citeauthor{Wang-23-03} recently built upon previous work to show that unsupervised denoising of non-equilibrium molecular structures leads to improved final NN potentials when fine-tuned on quantum-mechanical labels \cite{Sohl-Dickstein-15-11,Ho-20-12,Zaidi-22-10,Arts-23-02,Wang-23-03}.

Supervised pre-training tasks also exist. One technique of particular relevance here is ``domain knowledge injection''. \citeauthor{Shui-22-10} showed that learning to mimic existing empirical potentials can dramatically improve accuracies when fine-tuning on quantum-mechanical data \cite{Shui-22-10}. Our proposed technique builds upon this approach by improving the generality and accuracy of the knowledge that we inject during pre-training. 

\textbf{Transfer learning.}
Transfer learning involves using knowledge learned in one setting to improve performance on some other target task \cite{Zhuang-19-11}. Many strategies exist to perform this transfer of knowledge, and successful applications have been found in fields as varied as image classification \cite{Saenko-10} and captioning \cite{Vinyals-15-06}, gaming strategies \cite{Sharma-07-01}, and social network analysis \cite{Tang-16-04}.

In computational chemistry, transfer learning has been used to improve MLIP models \cite{Smith-19-07, Zhang-22-09, Chen-23-02} and molecular property prediction \cite{Li-22-10}. 
For example, \citeauthor{Smith-19-07} have shown transfer learning from DFT- to coupled-cluster- [CCSD(T)-] level data for molecules, using relatively few of the latter expensive labels to ``lift'' the level of the final MLIP \cite{Smith-19-07}.
We note that our approach can be recast in a similar light as transfer learning from MLIP- to DFT-level accuracy \cite{Gardner-22-12}.

\textbf{Alchemical learning.}
Alchemical learning (across chemical elements) seeks to improve model performance by training on data for elements with which the final model is not concerned.
In previous work \cite{Faber-18-03, Fias-19-01}, alchemical learning was performed by creating local-environment descriptors that explicitly include elemental information.
\citeauthor{Faber-18-03} \cite{Faber-18-03} and \citeauthor{Fias-19-01} \cite{Fias-19-01}, respectively, have shown that models incorporating such descriptors can (i) use training data including additional elements to improve their performance and (ii) extrapolate at test-time to elements not seen during training. 

In this work, we seek to learn chemically transferable model weights, implicitly creating features instead of manually crafting them. To this end, we pre-train NN potentials on synthetic data for carbon, and show that this improves performance when fine-tuning on DFT data for silicon.
We thus present alchemical pre-training as a form of domain adaptation transfer learning.

\section{Methodology}
\label{sec:methods}

\subsection{Neural-network interatomic potentials}

Neural networks are a powerful class of models used to construct ML potentials. 
Their main computational primitive, the multi-layer perceptron, can provably learn any function that maps from one fixed-dimensional space to another, given sufficient parameterisation \cite{Hornik-91-01}.
Much research effort has gone into creating atomic-environment features \cite{Behler-07-04,Bartok-13-05,Wang-18-07,Drautz-19-01} and model architectures \cite{Schuett-18-03,Unke-19-05,Ko-21-01,Schuett-21-06,Batzner-22-05,Batatia-22-10,Batatia-22-05,Simeon-23-06}, allowing one to use the fixed-dimensional mappings provided by MLPs to regress energies and forces onto atomic environments.
The compounding of these efforts over time has established NN potentials as an accurate and data-efficient approach in materials modelling.

Herein, we use the Neural Equivariant Interatomic Potentials (NequIP) architecture introduced by \citeauthor{Batzner-22-05} \cite{Batzner-22-05}. NequIP combines learnable embeddings with interaction blocks in a message passing framework. E3 equivariance is achieved by the use of geometric tensors as internal features: these are equivariant to rotation and reflection, a property that is conserved under the tensor products used within the interaction blocks. In the output layer, invariant, per-atom energies are predicted; forces are explicitly calculated as derivatives to ensure energy conservation \cite{Batzner-22-05}.

To train an NN model, a measure of its performance must be defined. Since NNs are fully differentiable, the gradient of this ``loss function'' can be calculated w.r.t.\ each parameter using backpropagation. Various optimisers exist that then use these gradients to update the values of parameters.
As per the NequIP paper, we define the loss for a single structure with energy label $E$ and force labels $\{F_j\}$ as:
$$
\mathcal{L} = \lambda_{\rm E} ||\hat{E} - E||^2 + \lambda_{\rm F} \frac{1}{3N}\sum_{j=1}^{N} \sum_{\alpha=1}^3 \left| \left| -\frac{\partial \hat{E}}{\partial r_{j, \alpha}} - F_{j, \alpha} \right| \right|^2
$$
where $\lambda_{\rm E}$ and $\lambda_{\rm F}$ weight the energy and force contributions to the total loss, respectively.
We compute several such losses on a mini-batch of structures before performing an optimisation step. The order that structures appear in these mini-batches is determined by the random seed used.

\subsection{Pre-training and fine-tuning}

\begin{figure*}[t]
    \centering
    \includegraphics[width=15cm]{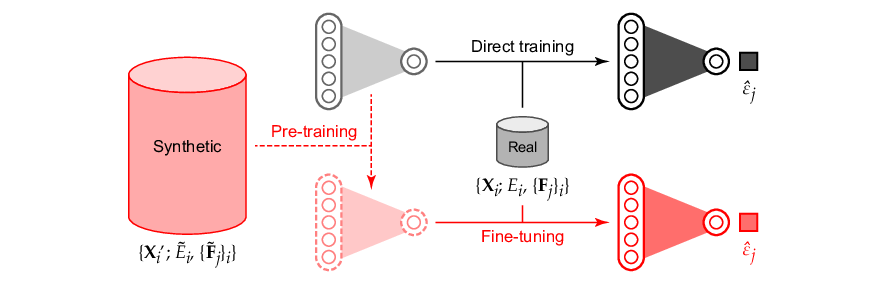}
    \caption{Synthetic pre-training for neural-network interatomic potentials. This schematic compares the established direct training approach ({\em top}) to synthetic pre-training and subsequent fine-tuning ({\em bottom}). Direct training starts from a randomly initialised NN model which is then optimised for real, quantum-mechanically computed energy and force data. Our approach, instead, involves an initial pre-training step, directly training a model on a very large, synthetic database generated by an existing ML potential model. The pre-trained model is then fine-tuned on the same real data as in the direct approach. (See also Ref.\ \citenum{Gardner-22-12}.)}
    \label{fig:schematic}
\end{figure*}

Various advanced methods exist to fine-tune a pre-trained model, including early layer freezing, AutoFreeze \cite{Liu-21-04}, application of the lottery ticket hypothesis \cite{Frankle-19-03}, and discriminative fine-tuning \cite{Howard-18-05}. Herein, we adopt the simplest possible approach and fine-tune all weights of the best pre-trained model by training as normal on the real data. Thus, apart from weight initialisation, our training procedures for direct training, pre-training, and fine-tuning are identical. An overview of the approach is given in \autoref{fig:schematic}.

\subsection{Datasets and synthetic labels}

The dataset of synthetic carbon structures used here, to which we refer as C-SYNTH-23M, is taken from Ref.~\citenum{Gardner-22-12}. The dataset contains 546 carbon molecular-dynamics (MD) trajectories for carbon structures with mass densities ranging from 1.0 to 3.5 g cm$^{-3}$. These trajectories were generated through LAMMPS \cite{Thompson-22-02} melt--quench--anneal simulations with varying parameters, driven by the C-GAP-17 ML potential \cite{Deringer-17-03} in the Gaussian Approximation Potential (GAP) framework \cite{Bartok-10-04}.
The dataset contains over 23 million unique atomic environments, including a variety of ordered and disordered local structures.
This dataset had originally been labelled with local energies and forces using C-GAP-17. We here re-label it with the following models: the C-GAP-20U potential \cite{Rowe-20-07}, the atomic cluster expansion (ACE) potential for carbon by \citeauthor{Qamar-23-06} \cite{Drautz-19-01,Qamar-23-06}, the long-range carbon bond order potential (LCBOP) \cite{Los-03-07}, and the environment-dependent interaction potential (EDIP) fitted for carbon \cite{Bazant-96-11, Marks-00-12}. 

We also used ``real'' (quantum-mechanically labelled) datasets. 
One of those was the C-GAP-17 training dataset \cite{Deringer-17-03}. This dataset contains cells representing liquid and amorphous carbon, an isolated dimer, as well as randomly distorted unit cells of graphite and diamond, labelled with atomic forces and local energies from DFT computations.
The C-GAP-20U training dataset was used to extract carbon nanotube structures \cite{Rowe-20-07}. 
The Si-GAP-18 training dataset was used to provide DFT-labelled silicon structures for alchemical learning experiments \cite{Bartok-18-12}.

\subsection{Computational details}

\textbf{Compute resources.} We used a single NVIDIA RTX A6000 GPU to train all NequIP models discussed in the present work, for a total of about 590 wallclock hours. Running at full power (300 W) this is 177 kWh, which in the UK is estimated as equivalent to about 37 kg CO$_2$.

\textbf{NequIP models.} We performed an initial sweep over salient model and training hyperparameters. Balancing speed and accuracy, we settled on 4 message-passing layers, each with 32 features and a message-passing cut-off of 4.0 \AA{}. The Adam optimiser \cite{Kingma-17-01} was used in training together with an exponential moving average decay rate of 0.99. Energy ({\tt PerAtomMSE} in eV/atom) and force ({\tt ForceMSE} in eV/\AA{}) losses were weighted 4:1 and all other parameters were kept as default.

\textbf{Software and data.} We used the NequIP library to train all models. The LCBOP \cite{Los-03-07, LCBOP-KIM-19} and EDIP \cite{EDIP-Kim-18} empirical force-field labels were generated using the {\tt openkim} library and models \cite{Tadmor-11,Elliott-11}. The {\tt pacemaker} \cite{Bochkarev-22,Lysogorskiy-21} and {\tt quippy} \cite{Csanyi-07,Kermode-20-03} Python packages were used to generate C-ACE and C-GAP-20U labels, respectively. LAMMPS was used to drive MD simulations \cite{Thompson-22-02}. Custom-written code, data, and configurations used to train the models in the present work are provided in a GitHub repository (see data availability statement below).

\section{Proof-of-concept}

\begin{figure*}[t]
    \centering
    \includegraphics[width=15cm]{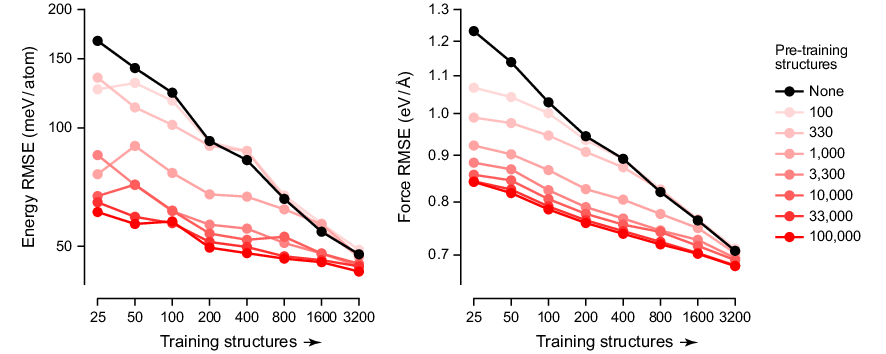}
    \caption{
    Proof-of-concept for synthetic pre-training. Energy ({\em left}) and force ({\em right}) RMSE as a function of the number of DFT labelled structures seen during training ($x$-axis) and number of synthetic pre-training structures (colour coding) as measured on the C-GAP-17 test set.
    Direct training points (black) are plotted as the best of 5 differently seeded runs. Fine-tuning points are plotted for a single random seed.
    }
    \label{fig:n-pre-train}
\end{figure*}

We show in \autoref{fig:n-pre-train} that synthetic pre-training on C-ACE \cite{Qamar-23-06} labels leads to more accurate MLIPs than the widely established direct training approach. Increasing the number of structures seen in pre-training systematically improves the performance of the fine-tuned model.

We directly trained a series of NequIP models on increasing amounts of the C-GAP-17 training set. We also synthetically pre-trained several models before fine-tuning on the same structures.
Analysing the results, we find gains in accuracy on the C-GAP-17 test set as a result of synthetic pre-training are largest when only small amounts of real data are available. Training on just 25 structures, we observe an improvement of 105 meV/atom (\textbf{63\%}) in per-atom energy RMSE, and $0.39$ eV/Å (\textbf{32\%}) in force component RMSE on the C-GAP-17 hold-out test set. We note that, to save time and compute, fine-tuning results are quoted for a single run, whereas the best model from 5 training runs is taken for direct training. Thus these quoted values provide a lower bound on the improvement that synthetic pre-training can bring.

Taking an orthogonal view, to achieve the same accuracy as a model directly trained on 800 structures, a synthetically pre-trained model requires no more than 25 structures. This amounts to a data efficiency saving of at least $32 \times$. 
This supports our original proposition in Ref. \citenum{Gardner-22-12} that this sort of pre-training is most relevant and useful when aiming to fine-tune on small numbers of high-level quantum-mechanical (beyond DFT) data.
Adopting the philosophy of pre-train once, fine-tune many times, we also find that the effective training time is dramatically reduced by synthetic pre-training: fine-tuning a model on 25 real structures took 35\,s on the GPU used. In contrast, direct training on 800 structures took between 15 and 44 minutes. This amounts to a wall-time speed-up of between $25\times$ and $75\times$ to achieve the same numerical accuracy.

\section{Experiments}

\subsection{The synthetic source matters}

\begin{figure*}[t]
    \centering
    \includegraphics[width=15cm]{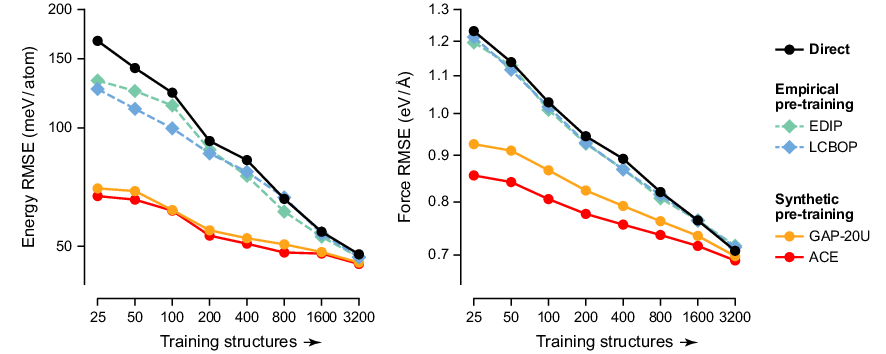}
    \caption{
    Effect of using different sources for pre-training. Per-atom energy ({\em left}) and component-wise force ({\em right}) RMSE are plotted as a function of the number of real structures seen during training, as measured on the C-GAP-17 hold-out test set \cite{Deringer-17-03}. Pre-training on empirical force-field labels leads to limited improvement over direct-training, whereas pre-training on existing MLP labels leads to significant improvements.
    We pre-train each model on the same 10,000 synthetic structures taken at random from C-SYNTH-23M, and report the best error from 5 direct-training or fine-tuning attempts.
    }
    \label{fig:pre-train-sources}
\end{figure*}

We studied the effect of the source of the synthetic labels -- that is, at what level are the pre-training data computed?
Previous work has shown that empirical force fields can act as a source of synthetic labels for pre-training, improving energy errors by more than $50\%$ in some settings \cite{Shui-22-10}.
We carried out similar experiments for the specific dataset (C-GAP-17) and model architecture (NequIP) used herein, making use of two popular empirical force fields for carbon, {\em viz.}\ EDIP and LCBOP.
Learning curves are presented in \autoref{fig:pre-train-sources} (showing the best of 5 direct-training or fine-tuning attempts in each case), and further data may be found in \autoref{tab:synthetic-source-errors} (including standard deviations).

We find that, in the low-data limit (25 DFT labelled structures), empirical pre-training leads to a $25\%$ improvement in energy RMSE. However, the absolute values are still well above ``chemical accuracy'' ($\sim 40$ meV/atom), and the force errors improve at most by $3\%$ compared to the directly trained models. 
In contrast, we find that pre-training on C-GAP-20U or C-ACE labels leads to larger improvements in energy and force accuracy, and in data efficiency, than pre-training with empirical interatomic potentials.
One explanation for this observed behaviour could be that the latter potentials have been parameterised to describe limited regions of configurational space. 
Indeed, their predictions on the test set as a whole appear to be meaningless (force RMSEs $>5$ eV/atom). 
Thus, the knowledge injected into the NequIP model during pre-training is only useful for comparatively fewer structures.
In contrast, the C-ACE and C-GAP-20U models aim to describe, at comparatively larger computational cost, much larger regions of configurational space, and so pre-training on them helps to bring down the error across the whole test set.

\begin{table}
\small
\centering
\setlength\tabcolsep{4pt}
\caption{Per-atom energy RMSE (meV/atom) as a function of the pre-training source and number of DFT-labelled training structures. Mean values are provided together with standard deviations over 5 separate fine-tuning runs for each pre-trained model. Values marked with a $^*$ indicate large standard deviations -- in these cases, one or more of the direct-training or fine-tuning runs experienced a sudden increase in loss and failed to converge.}
\label{tab:synthetic-source-errors}
\begin{tabular*}{\textwidth}{
  @{\extracolsep{\fill}}
  c
  c|
  c|
  c
  c
  c
  c
}
\hline
 &  &   & \multicolumn{4}{c}{Pre-training label source}\\
 &  & {Direct} & {EDIP \cite{Marks-00-12}} & {LCBOP \cite{Los-03-07}} & {GAP-20U \cite{Rowe-20-07}} & {C-ACE \cite{Qamar-23-06}} \\
\hline
\hline
\multirow{8}{*}{\begin{tabular}[c]{@{}c@{}}Training\\structures\end{tabular}} 
& 25 & 192.3 $\pm$ 21.2 & 136.4 $\pm$ 5.8 & 129.9 $\pm$ 3.9 & 73.1 $\pm$ 3.6 & 70.3 $\pm$ 2.8 \\
& 50 & 166.1 $\pm$ 15.5 & 134.6 $\pm$ 7.1 & 119.0 $\pm$ 4.7 & 69.7 $\pm$ 1.2 & 71.8 $\pm$ 6.9 \\
& 100 & 129.2 $\pm$ 5.3 & 115.6 $\pm$ 1.4 & 106.3 $\pm$ 6.4 & 63.1 $\pm$ 1.1 & 63.0 $\pm$ 1.5 \\
& 200 & 96.1 $\pm$ 2.8 & 90.9 $\pm$ 2.7 & 87.1 $\pm$ 0.8 & 55.5 $\pm$ 0.4 & 53.8 $\pm$ 0.6 \\
& 400 & 87.2 $\pm$ 3.5 & 83.3 $\pm$ 6.2 & 80.6 $\pm$ 2.3 & 52.8 $\pm$ 0.6 & 51.9 $\pm$ 1.0 \\
& 800 & 69.1 $\pm$ 2.8 & 71.1 $\pm$ 19.2$^*$ & 68.1 $\pm$ 1.3 & 51.4 $\pm$ 0.8 & 50.2 $\pm$ 1.7 \\
& 1600 & 56.0 $\pm$ 1.5 & 56.1 $\pm$ 4.0 & 54.9 $\pm$ 0.8 & 49.2 $\pm$ 0.8 & 48.2 $\pm$ 0.4 \\
& 3200 & 51.0 $\pm$ 6.2$^*$ & 88.2 $\pm$ 72.3$^*$ & 48.2 $\pm$ 0.9 & 45.9 $\pm$ 0.5 & 45.4 $\pm$ 0.4 \\
\hline
\end{tabular*}
\end{table}

\subsection{Pre-training can improve stability}

\begin{figure}[!t]
    \centering
    \includegraphics[width=15cm]{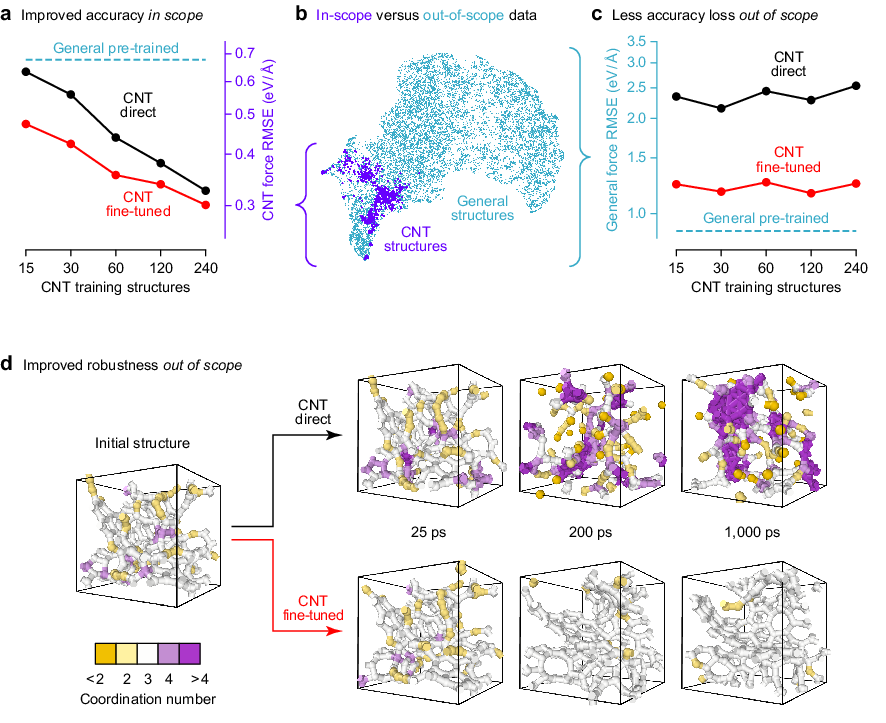}
    \caption{ Training directly on carbon nanotube (CNT) structures leads to potentials with very limited physical understanding. Fine-tuning from a potential pre-trained on a general carbon dataset leads to much better performance. 
    (\textbf{a}) RMSE on a CNT test set for direct, pre-trained and fine-tuned models. The fine-tuned model has significantly better numerical performance.
    (\textbf{b}) Projection of a random sample of carbon nanotube environments onto an embedding of the chemical space spanned by the general pre-training set (created by describing each atomic environment using a SOAP vector \cite{Bartok-13-05} and performing dimensionality reduction using UMAP) \cite{McInnes-20-09}.
    (\textbf{c}) RMSE on a general-purpose test set. The direct model performs very poorly, which is to be expected as it has not encountered environments other than CNT structures during training; the fine-tuning process leads to some degradation in the fine-tuned models' general performance. In panels (a) and (c), we plot results for the best model taken from 5 differently seeded training runs for the direct and fine-tuned models.
    (\textbf{d}) Visualisation of relevant structures from MD trajectories driven by a model directly trained on the CNT data only ({\em top}) and a fine-tuned model ({\em bottom}).
    Structural images were created using OVITO \cite{OVITO}.
    }
    \label{fig:bucky-md}
\end{figure}

Chemists using MLIPs for research are often interested in, and have small amounts of accurately labelled data for, highly specialised regions of chemical space.
Fitting models directly to such restricted datasets can lead to models with poor quantitative accuracy and unphysical behaviour outside the scope of their training data.
We propose general synthetic pre-training as a means to bypass both of these issues.

To demonstrate this, we sourced a set of structures consisting exclusively of carbon nanotubes (CNT), previously labelled with DFT, from the C-GAP-20U dataset \cite{Rowe-20-07}.
We take a generally pre-trained model, {\emph i.e.}, a model pre-trained on C-ACE labels for 10,000 carbon structures covering a wide range of densities and degrees of structural disorder taken from C-SYNTH-23M, and we compare fine-tuning from this to directly training on the small, specialised dataset (\autoref{fig:bucky-md}a).
As above, we see that synthetic pre-training significantly improves test-set accuracy (here $\sim$25\% improvement in force RMSE). 

In \autoref{fig:bucky-md}c we compare force errors as measured on the general C-GAP-17 test set. 
Directly trained models, having never seen a vast majority of this configurational space, have no reason to perform well here, and indeed they do not.
In contrast, the pre-trained model was trained on lots of data covering a vast majority of the chemical space spanned by this test set. It thus performs very well, approaching the level of the pre-training source on this dataset. 
The fine-tuning process leads to a deterioration of this general performance: it seems that the penalty for improving on the CNT structures is partial forgetting of the original pre-training data.
However, compared to the directly trained models, the fine-tuned models still perform significantly better (only 1.4$\times$ the error of the pre-trained model, as compared to 2.7$\times$). 
Thus synthetic pre-training leads to models that are more robust when making predictions on atomic environments that are not present in the final training set.

To further emphasise this robustness, we perform MD simulations using the direct and fine-tuned models (\autoref{fig:bucky-md}d). Taking an amorphous (1.5 g\,cm$^{-3}$) carbon structure, we heat from 300\,K to 3,000\,K over 100\,ps, before holding for a cumulative 1,000\,ps.
We find that the fine-tuned model creates stable trajectories that capture the qualitatively correct behaviour, {\em viz.}\ graphitisation \cite{Powles-09-02, DeTomas-16-11}.
In contrast, the directly-trained models fail rapidly and dramatically by failing to handle rare events. A brief period ($\sim$20\,ps) of reasonable dynamics is interrupted by the creation of a 5-fold-connected carbon atom. After this, the simulation rapidly breaks down, leading to simultaneous production of 0-coordinated free carbon atoms and clusters of highly (8+) coordinated atoms.

\subsection{Alchemical pre-training can be useful}

\begin{figure}[!t]
    \centering
    \includegraphics[width=15cm]{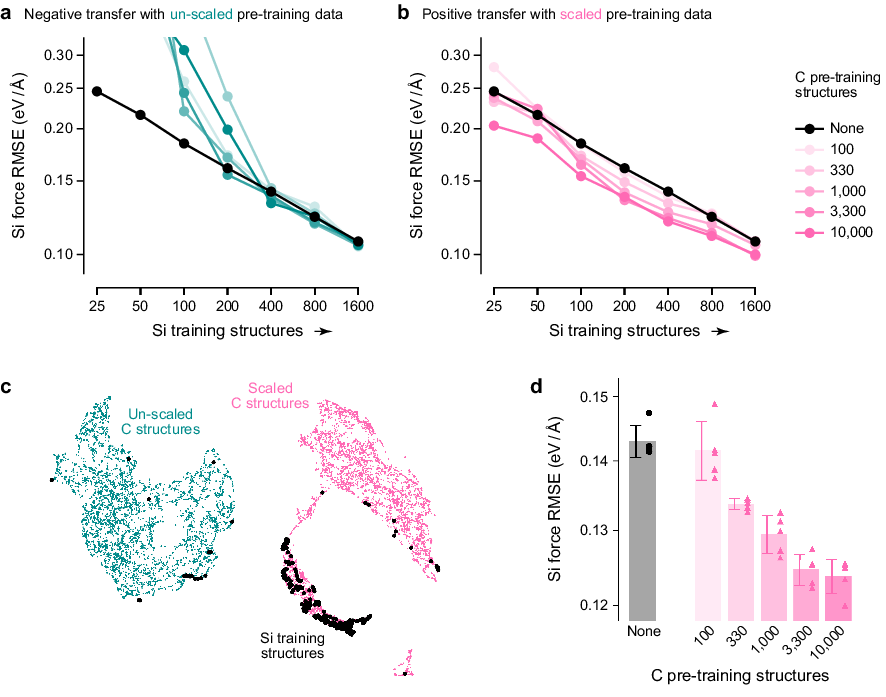}
    \caption{Alchemical pre-training. (\textbf{a}) Pre-training to mimic C-ACE labels on unscaled carbon structures (turquoise) leads to, at best, no significant improvement upon fine-tuning. (\textbf{b}) Scaling carbon structures by $1.5\times$ (magenta) leads to successful transfer. In both sets of learning curves, we plot the best model performance as taken over 5 separate fine-tuning attempts. (\textbf{c}) UMAP projections of the Si dataset (black) onto the structural space spanned by the un-/scaled structures show that the sp$^3$-like region of the scaled carbon dataset overlaps well with the majority of the silicon structures. (\textbf{d}) Mean Si test-set force RMSE as taken over 5 separate training runs as a function of the number of scaled C structures seen during pre-training before fine-tuning on 400 Si structures.}
    \label{fig:alchemy}
\end{figure}

We have shown that synthetic pre-training is useful when a general MLIP has already been trained for the system of interest. 
However, when exploring a new chemical system, this will often not be the case.
We therefore explore using an MLIP trained on a different chemical system as a source for \emph{alchemical} synthetic pre-training data (\autoref{fig:alchemy}).

We find that using the carbon structures and labels as they are yields no improvements, and in the low-data limit leads to significantly negative transfer. We note, however, that the characteristic length-scales of carbon and silicon structures are very different: the typical minimum separation seen in carbon structures is $\sim$1.4 \AA{} whereas for silicon it is $\sim$2.1 \AA. We therefore expand all carbon structures by a factor of $2.1/1.4=1.5$, keeping the labels as they are, and repeat the experiment by pre-training with the scaled structures and fine-tuning on the same silicon dataset. In this case, we observe positive transfer, with a similar trend in the number of pre-training structures as seen previously.

\autoref{fig:alchemy}c illustrates the effect of this scaling procedure. We project environments from the Si dataset onto a UMAP embedding \cite{McInnes-20-09} of environments from the unscaled (turquoise) and scaled (magenta) carbon structures. 
A majority of Si structures lie in the ``sp$^3$-like'', tetrahedral region of the scaled carbon structures -- we posit that it is the knowledge of this region gained during general pre-training that is leading to the positive transfer. Hence, we believe that a current limitation of this technique is that the alchemical structures used for pre-training appear to have to share similarities with the fine-tuning target. Further work on this theme is ongoing.

\section{Conclusions}
\label{sec:conclusions}

We have shown that pre-training neural-network interatomic potentials on synthetic energy and force data can improve accuracy, training time, and data efficiency for end users. We have focused on a single architecture (NequIP) and chemical system (carbon), but we expect that the technique in principle is more general, which we will explore in future work.

We presented a series of experiments to test the capabilities (and limits) of the approach. We found that increasing the number of structures seen during pre-training improves the final model, but that these improvements decay rapidly in the limit of large fine-tuning data (\autoref{fig:n-pre-train}), consistent with our previous findings for regressing scalar quantities \cite{Gardner-22-12}.
We showed that the synthetic source matters -- the more accurate the pre-training labels, the larger the gain from the approach -- and that synthetic pre-training can improve robustness of NN potentials in principle.

A possible limitation of homogenous synthetic pre-training is that one requires an existing MLIP to have been trained on this particular system -- this is readily available for carbon, but will not always be the case for more complex chemistries. In an effort to bypass this restriction, we showed a proof-of-concept for alchemical pre-training: using an MLIP trained on one chemical system (here, carbon) to pre-train a model for use on a different system (here, silicon). We observe positive transfer, and further work is ongoing to improve the magnitude of this improvement.

\section*{Data availability}

Data, code, and model weights supporting the present work are available at\\
\url{https://github.com/jla-gardner/nnp-pre-training}.

\section*{Acknowledgements}

We thank Z. El-Machachi, Z. Faure Beaulieu, and D. F. Thomas du Toit for discussions and for help with data labelling. 
J.L.A.G. acknowledges a UKRI Linacre - The EPA Cephalosporin Scholarship, support from an EPSRC DTP award [grant number EP/T517811/1], and from the Department of Chemistry, University of Oxford.
V.L.D. acknowledges a UK Research and Innovation Frontier Research grant [grant number EP/X016188/1].

\bibliographystyle{apsrev}

\begin{thebibliography}{74}
\expandafter\ifx\csname natexlab\endcsname\relax\def\natexlab#1{#1}\fi
\expandafter\ifx\csname bibnamefont\endcsname\relax
  \def\bibnamefont#1{#1}\fi
\expandafter\ifx\csname bibfnamefont\endcsname\relax
  \def\bibfnamefont#1{#1}\fi
\expandafter\ifx\csname citenamefont\endcsname\relax
  \def\citenamefont#1{#1}\fi
\expandafter\ifx\csname url\endcsname\relax
  \def\url#1{\texttt{#1}}\fi
\expandafter\ifx\csname urlprefix\endcsname\relax\def\urlprefix{URL }\fi
\providecommand{\bibinfo}[2]{#2}
\providecommand{\eprint}[2][]{\url{#2}}

\bibitem[{\citenamefont{Behler}(2017)}]{Behler-17-05}
\bibinfo{author}{\bibfnamefont{J.}~\bibnamefont{Behler}},
  \bibinfo{journal}{Angew. Chem. Int. Ed.} \textbf{\bibinfo{volume}{56}},
  \bibinfo{pages}{12828} (\bibinfo{year}{2017}).

\bibitem[{\citenamefont{Deringer et~al.}(2019)\citenamefont{Deringer, Caro, and
  Cs\'a{}nyi}}]{Deringer-19-09}
\bibinfo{author}{\bibfnamefont{V.~L.} \bibnamefont{Deringer}},
  \bibinfo{author}{\bibfnamefont{M.~A.} \bibnamefont{Caro}}, \bibnamefont{and}
  \bibinfo{author}{\bibfnamefont{G.}~\bibnamefont{Cs\'a{}nyi}},
  \bibinfo{journal}{Adv. Mater.} \textbf{\bibinfo{volume}{31}},
  \bibinfo{pages}{1902765} (\bibinfo{year}{2019}).

\bibitem[{\citenamefont{No\'e{} et~al.}(2020)\citenamefont{No\'e{}, Tkatchenko,
  M\"u{}ller, and Clementi}}]{Noe-20-02}
\bibinfo{author}{\bibfnamefont{F.}~\bibnamefont{No\'e{}}},
  \bibinfo{author}{\bibfnamefont{A.}~\bibnamefont{Tkatchenko}},
  \bibinfo{author}{\bibfnamefont{K.-R.} \bibnamefont{M\"u{}ller}},
  \bibnamefont{and} \bibinfo{author}{\bibfnamefont{C.}~\bibnamefont{Clementi}},
  \bibinfo{journal}{Annu. Rev. Phys. Chem.} \textbf{\bibinfo{volume}{71}},
  \bibinfo{pages}{361} (\bibinfo{year}{2020}).

\bibitem[{\citenamefont{Unke et~al.}(2021)\citenamefont{Unke, Chmiela, Sauceda,
  Gastegger, Poltavsky, Sch\"u{}tt, Tkatchenko, and M\"u{}ller}}]{Unke-21-03}
\bibinfo{author}{\bibfnamefont{O.~T.} \bibnamefont{Unke}},
  \bibinfo{author}{\bibfnamefont{S.}~\bibnamefont{Chmiela}},
  \bibinfo{author}{\bibfnamefont{H.~E.} \bibnamefont{Sauceda}},
  \bibinfo{author}{\bibfnamefont{M.}~\bibnamefont{Gastegger}},
  \bibinfo{author}{\bibfnamefont{I.}~\bibnamefont{Poltavsky}},
  \bibinfo{author}{\bibfnamefont{K.~T.} \bibnamefont{Sch\"u{}tt}},
  \bibinfo{author}{\bibfnamefont{A.}~\bibnamefont{Tkatchenko}},
  \bibnamefont{and} \bibinfo{author}{\bibfnamefont{K.-R.}
  \bibnamefont{M\"u{}ller}}, \bibinfo{journal}{Chem. Rev.}
  \textbf{\bibinfo{volume}{121}}, \bibinfo{pages}{10142}
  (\bibinfo{year}{2021}).

\bibitem[{\citenamefont{Friederich et~al.}(2021)\citenamefont{Friederich,
  H\"a{}se, Proppe, and Aspuru-Guzik}}]{Friederich-21-05}
\bibinfo{author}{\bibfnamefont{P.}~\bibnamefont{Friederich}},
  \bibinfo{author}{\bibfnamefont{F.}~\bibnamefont{H\"a{}se}},
  \bibinfo{author}{\bibfnamefont{J.}~\bibnamefont{Proppe}}, \bibnamefont{and}
  \bibinfo{author}{\bibfnamefont{A.}~\bibnamefont{Aspuru-Guzik}},
  \bibinfo{journal}{Nat. Mater.} \textbf{\bibinfo{volume}{20}},
  \bibinfo{pages}{750} (\bibinfo{year}{2021}).

\bibitem[{\citenamefont{Cheng et~al.}(2020)\citenamefont{Cheng, Mazzola,
  Pickard, and Ceriotti}}]{Cheng-20-09}
\bibinfo{author}{\bibfnamefont{B.}~\bibnamefont{Cheng}},
  \bibinfo{author}{\bibfnamefont{G.}~\bibnamefont{Mazzola}},
  \bibinfo{author}{\bibfnamefont{C.~J.} \bibnamefont{Pickard}},
  \bibnamefont{and} \bibinfo{author}{\bibfnamefont{M.}~\bibnamefont{Ceriotti}},
  \bibinfo{journal}{Nature} \textbf{\bibinfo{volume}{585}},
  \bibinfo{pages}{217} (\bibinfo{year}{2020}).

\bibitem[{\citenamefont{Zhou et~al.}(2022)\citenamefont{Zhou, Kirkpatrick, and
  Deringer}}]{Zhou-22-02}
\bibinfo{author}{\bibfnamefont{Y.}~\bibnamefont{Zhou}},
  \bibinfo{author}{\bibfnamefont{W.}~\bibnamefont{Kirkpatrick}},
  \bibnamefont{and} \bibinfo{author}{\bibfnamefont{V.~L.}
  \bibnamefont{Deringer}}, \bibinfo{journal}{Adv. Mater.}
  \textbf{\bibinfo{volume}{34}}, \bibinfo{pages}{2107515}
  (\bibinfo{year}{2022}).

\bibitem[{\citenamefont{Westermayr et~al.}(2022)\citenamefont{Westermayr,
  Gastegger, V\"o{}r\"o{}s, Panzenboeck, Joerg, Gonz\'a{}lez, and
  Marquetand}}]{Westermayr-22-08}
\bibinfo{author}{\bibfnamefont{J.}~\bibnamefont{Westermayr}},
  \bibinfo{author}{\bibfnamefont{M.}~\bibnamefont{Gastegger}},
  \bibinfo{author}{\bibfnamefont{D.}~\bibnamefont{V\"o{}r\"o{}s}},
  \bibinfo{author}{\bibfnamefont{L.}~\bibnamefont{Panzenboeck}},
  \bibinfo{author}{\bibfnamefont{F.}~\bibnamefont{Joerg}},
  \bibinfo{author}{\bibfnamefont{L.}~\bibnamefont{Gonz\'a{}lez}},
  \bibnamefont{and}
  \bibinfo{author}{\bibfnamefont{P.}~\bibnamefont{Marquetand}},
  \bibinfo{journal}{Nat. Chem.} \textbf{\bibinfo{volume}{14}},
  \bibinfo{pages}{914} (\bibinfo{year}{2022}).

\bibitem[{\citenamefont{Azizi et~al.}(2023)\citenamefont{Azizi, Kornblith,
  Saharia, Norouzi, and Fleet}}]{Azizi-23-04}
\bibinfo{author}{\bibfnamefont{S.}~\bibnamefont{Azizi}},
  \bibinfo{author}{\bibfnamefont{S.}~\bibnamefont{Kornblith}},
  \bibinfo{author}{\bibfnamefont{C.}~\bibnamefont{Saharia}},
  \bibinfo{author}{\bibfnamefont{M.}~\bibnamefont{Norouzi}}, \bibnamefont{and}
  \bibinfo{author}{\bibfnamefont{D.~J.} \bibnamefont{Fleet}},
  \emph{\bibinfo{title}{Synthetic {{Data}} from {{Diffusion Models Improves
  ImageNet Classification}}}} (\bibinfo{year}{2023}),
  \eprint{arXiv:2304.08466}.

\bibitem[{\citenamefont{Kirillov et~al.}(2023)\citenamefont{Kirillov, Mintun,
  Ravi, Mao, Rolland, Gustafson, Xiao, Whitehead, Berg, Lo
  et~al.}}]{Kirillov-23-04}
\bibinfo{author}{\bibfnamefont{A.}~\bibnamefont{Kirillov}},
  \bibinfo{author}{\bibfnamefont{E.}~\bibnamefont{Mintun}},
  \bibinfo{author}{\bibfnamefont{N.}~\bibnamefont{Ravi}},
  \bibinfo{author}{\bibfnamefont{H.}~\bibnamefont{Mao}},
  \bibinfo{author}{\bibfnamefont{C.}~\bibnamefont{Rolland}},
  \bibinfo{author}{\bibfnamefont{L.}~\bibnamefont{Gustafson}},
  \bibinfo{author}{\bibfnamefont{T.}~\bibnamefont{Xiao}},
  \bibinfo{author}{\bibfnamefont{S.}~\bibnamefont{Whitehead}},
  \bibinfo{author}{\bibfnamefont{A.~C.} \bibnamefont{Berg}},
  \bibinfo{author}{\bibfnamefont{W.-Y.} \bibnamefont{Lo}},
  \bibnamefont{et~al.}, \emph{\bibinfo{title}{Segment {{Anything}}}}
  (\bibinfo{year}{2023}), \eprint{arXiv:2304.02643}.

\bibitem[{\citenamefont{To et~al.}(2023)\citenamefont{To, Bui, Guo, and
  Nguyen}}]{To-23-05}
\bibinfo{author}{\bibfnamefont{H.~Q.} \bibnamefont{To}},
  \bibinfo{author}{\bibfnamefont{N.~D.~Q.} \bibnamefont{Bui}},
  \bibinfo{author}{\bibfnamefont{J.}~\bibnamefont{Guo}}, \bibnamefont{and}
  \bibinfo{author}{\bibfnamefont{T.~N.} \bibnamefont{Nguyen}},
  \emph{\bibinfo{title}{Better {{Language Models}} of {{Code}} through
  {{Self-Improvement}}}} (\bibinfo{year}{2023}), \eprint{arXiv:2304.01228}.

\bibitem[{\citenamefont{Zhang et~al.}(2022{\natexlab{a}})\citenamefont{Zhang,
  Qin, Park, Han, Chiu, Pang, Le, and Wu}}]{Zhang-22-07}
\bibinfo{author}{\bibfnamefont{Y.}~\bibnamefont{Zhang}},
  \bibinfo{author}{\bibfnamefont{J.}~\bibnamefont{Qin}},
  \bibinfo{author}{\bibfnamefont{D.~S.} \bibnamefont{Park}},
  \bibinfo{author}{\bibfnamefont{W.}~\bibnamefont{Han}},
  \bibinfo{author}{\bibfnamefont{C.-C.} \bibnamefont{Chiu}},
  \bibinfo{author}{\bibfnamefont{R.}~\bibnamefont{Pang}},
  \bibinfo{author}{\bibfnamefont{Q.~V.} \bibnamefont{Le}}, \bibnamefont{and}
  \bibinfo{author}{\bibfnamefont{Y.}~\bibnamefont{Wu}},
  \emph{\bibinfo{title}{Pushing the {{Limits}} of {{Semi-Supervised Learning}}
  for {{Automatic Speech Recognition}}}} (\bibinfo{year}{2022}{\natexlab{a}}),
  \eprint{arXiv:2010.10504}.

\bibitem[{\citenamefont{Aty et~al.}(2022)\citenamefont{Aty, Strutt, Mcintyre,
  Allen, Barlow, {P{\'a}ez-P{\'e}rez}, Seddon, Brooks, Ces, and
  Gould}}]{Aty-22-04}
\bibinfo{author}{\bibfnamefont{H.~A.} \bibnamefont{Aty}},
  \bibinfo{author}{\bibfnamefont{R.}~\bibnamefont{Strutt}},
  \bibinfo{author}{\bibfnamefont{N.}~\bibnamefont{Mcintyre}},
  \bibinfo{author}{\bibfnamefont{M.}~\bibnamefont{Allen}},
  \bibinfo{author}{\bibfnamefont{N.~E.} \bibnamefont{Barlow}},
  \bibinfo{author}{\bibfnamefont{M.}~\bibnamefont{{P{\'a}ez-P{\'e}rez}}},
  \bibinfo{author}{\bibfnamefont{J.~M.} \bibnamefont{Seddon}},
  \bibinfo{author}{\bibfnamefont{N.}~\bibnamefont{Brooks}},
  \bibinfo{author}{\bibfnamefont{O.}~\bibnamefont{Ces}}, \bibnamefont{and}
  \bibinfo{author}{\bibfnamefont{I.~R.} \bibnamefont{Gould}},
  \bibinfo{journal}{Digital Discovery} \textbf{\bibinfo{volume}{1}},
  \bibinfo{pages}{98} (\bibinfo{year}{2022}).

\bibitem[{\citenamefont{Anker et~al.}(2023)\citenamefont{Anker, Butler, Le,
  Perring, and Thiyagalingam}}]{Anker-23-03}
\bibinfo{author}{\bibfnamefont{A.~S.} \bibnamefont{Anker}},
  \bibinfo{author}{\bibfnamefont{K.~T.} \bibnamefont{Butler}},
  \bibinfo{author}{\bibfnamefont{M.~D.} \bibnamefont{Le}},
  \bibinfo{author}{\bibfnamefont{T.~G.} \bibnamefont{Perring}},
  \bibnamefont{and}
  \bibinfo{author}{\bibfnamefont{J.}~\bibnamefont{Thiyagalingam}},
  \bibinfo{journal}{Digital Discovery} \textbf{\bibinfo{volume}{2}},
  \bibinfo{pages}{578} (\bibinfo{year}{2023}).

\bibitem[{\citenamefont{Schuetzke et~al.}(2023)\citenamefont{Schuetzke,
  Szymanski, and Reischl}}]{Schuetzke-23-06}
\bibinfo{author}{\bibfnamefont{J.}~\bibnamefont{Schuetzke}},
  \bibinfo{author}{\bibfnamefont{N.~J.} \bibnamefont{Szymanski}},
  \bibnamefont{and} \bibinfo{author}{\bibfnamefont{M.}~\bibnamefont{Reischl}},
  \bibinfo{journal}{npj Comput. Mater.} \textbf{\bibinfo{volume}{9}},
  \bibinfo{pages}{100} (\bibinfo{year}{2023}).

\bibitem[{\citenamefont{Morrow and Deringer}(2022)}]{Morrow-22-09}
\bibinfo{author}{\bibfnamefont{J.~D.} \bibnamefont{Morrow}} \bibnamefont{and}
  \bibinfo{author}{\bibfnamefont{V.~L.} \bibnamefont{Deringer}},
  \bibinfo{journal}{J. Chem. Phys.} \textbf{\bibinfo{volume}{157}},
  \bibinfo{pages}{104105} (\bibinfo{year}{2022}).

\bibitem[{\citenamefont{Gardner et~al.}(2022)\citenamefont{Gardner, {Faure
  Beaulieu}, and Deringer}}]{Gardner-22-12}
\bibinfo{author}{\bibfnamefont{J.~L.~A.} \bibnamefont{Gardner}},
  \bibinfo{author}{\bibfnamefont{Z.}~\bibnamefont{{Faure Beaulieu}}},
  \bibnamefont{and} \bibinfo{author}{\bibfnamefont{V.~L.}
  \bibnamefont{Deringer}}, \bibinfo{journal}{Digital Discovery}
  \textbf{\bibinfo{volume}{2}}, \bibinfo{pages}{651} (\bibinfo{year}{2022}).

\bibitem[{\citenamefont{{Faure Beaulieu} et~al.}(2023)\citenamefont{{Faure
  Beaulieu}, Nicholas, Gardner, Goodwin, and Deringer}}]{FaureBeaulieu-23-05}
\bibinfo{author}{\bibfnamefont{Z.}~\bibnamefont{{Faure Beaulieu}}},
  \bibinfo{author}{\bibfnamefont{T.~C.} \bibnamefont{Nicholas}},
  \bibinfo{author}{\bibfnamefont{J.~L.~A.} \bibnamefont{Gardner}},
  \bibinfo{author}{\bibfnamefont{A.~L.} \bibnamefont{Goodwin}},
  \bibnamefont{and} \bibinfo{author}{\bibfnamefont{V.~L.}
  \bibnamefont{Deringer}} (\bibinfo{year}{2023}), \eprint{arXiv:2305.05536}.

\bibitem[{\citenamefont{Kelvinius et~al.}(2023)\citenamefont{Kelvinius,
  Georgiev, Toshev, and Gasteiger}}]{Kelvinius-23-06}
\bibinfo{author}{\bibfnamefont{F.~E.} \bibnamefont{Kelvinius}},
  \bibinfo{author}{\bibfnamefont{D.}~\bibnamefont{Georgiev}},
  \bibinfo{author}{\bibfnamefont{A.~P.} \bibnamefont{Toshev}},
  \bibnamefont{and}
  \bibinfo{author}{\bibfnamefont{J.}~\bibnamefont{Gasteiger}},
  \emph{\bibinfo{title}{Accelerating {{Molecular Graph Neural Networks}} via
  {{Knowledge Distillation}}}} (\bibinfo{year}{2023}),
  \eprint{arXiv:2306.14818}.

\bibitem[{\citenamefont{Wang et~al.}(2023)\citenamefont{Wang, Xu, Li, and
  Farimani}}]{Wang-23-03}
\bibinfo{author}{\bibfnamefont{Y.}~\bibnamefont{Wang}},
  \bibinfo{author}{\bibfnamefont{C.}~\bibnamefont{Xu}},
  \bibinfo{author}{\bibfnamefont{Z.}~\bibnamefont{Li}}, \bibnamefont{and}
  \bibinfo{author}{\bibfnamefont{A.~B.} \bibnamefont{Farimani}},
  \emph{\bibinfo{title}{Denoise {{Pre-training}} on {{Non-equilibrium
  Molecules}} for {{Accurate}} and {{Transferable Neural Potentials}}}}
  (\bibinfo{year}{2023}), \eprint{arXiv:2303.02216}.

\bibitem[{\citenamefont{{Sohl-Dickstein}
  et~al.}(2015)\citenamefont{{Sohl-Dickstein}, Weiss, Maheswaranathan, and
  Ganguli}}]{Sohl-Dickstein-15-11}
\bibinfo{author}{\bibfnamefont{J.}~\bibnamefont{{Sohl-Dickstein}}},
  \bibinfo{author}{\bibfnamefont{E.~A.} \bibnamefont{Weiss}},
  \bibinfo{author}{\bibfnamefont{N.}~\bibnamefont{Maheswaranathan}},
  \bibnamefont{and} \bibinfo{author}{\bibfnamefont{S.}~\bibnamefont{Ganguli}},
  \emph{\bibinfo{title}{Deep {{Unsupervised Learning}} using {{Nonequilibrium
  Thermodynamics}}}} (\bibinfo{year}{2015}), \eprint{arXiv:1503.03585}.

\bibitem[{\citenamefont{Ho et~al.}(2020)\citenamefont{Ho, Jain, and
  Abbeel}}]{Ho-20-12}
\bibinfo{author}{\bibfnamefont{J.}~\bibnamefont{Ho}},
  \bibinfo{author}{\bibfnamefont{A.}~\bibnamefont{Jain}}, \bibnamefont{and}
  \bibinfo{author}{\bibfnamefont{P.}~\bibnamefont{Abbeel}},
  \emph{\bibinfo{title}{Denoising {{Diffusion Probabilistic Models}}}}
  (\bibinfo{year}{2020}), \eprint{arXiv:2006.11239}.

\bibitem[{\citenamefont{Zaidi et~al.}(2022)\citenamefont{Zaidi, Schaarschmidt,
  Martens, Kim, Teh, {Sanchez-Gonzalez}, Battaglia, Pascanu, and
  Godwin}}]{Zaidi-22-10}
\bibinfo{author}{\bibfnamefont{S.}~\bibnamefont{Zaidi}},
  \bibinfo{author}{\bibfnamefont{M.}~\bibnamefont{Schaarschmidt}},
  \bibinfo{author}{\bibfnamefont{J.}~\bibnamefont{Martens}},
  \bibinfo{author}{\bibfnamefont{H.}~\bibnamefont{Kim}},
  \bibinfo{author}{\bibfnamefont{Y.~W.} \bibnamefont{Teh}},
  \bibinfo{author}{\bibfnamefont{A.}~\bibnamefont{{Sanchez-Gonzalez}}},
  \bibinfo{author}{\bibfnamefont{P.}~\bibnamefont{Battaglia}},
  \bibinfo{author}{\bibfnamefont{R.}~\bibnamefont{Pascanu}}, \bibnamefont{and}
  \bibinfo{author}{\bibfnamefont{J.}~\bibnamefont{Godwin}},
  \emph{\bibinfo{title}{Pre-training via {{Denoising}} for {{Molecular Property
  Prediction}}}} (\bibinfo{year}{2022}), \eprint{arXiv:2206.00133}.

\bibitem[{\citenamefont{Arts et~al.}(2023)\citenamefont{Arts, Satorras, Huang,
  Zuegner, Federici, Clementi, No{\'e}, Pinsler, and van~den
  Berg}}]{Arts-23-02}
\bibinfo{author}{\bibfnamefont{M.}~\bibnamefont{Arts}},
  \bibinfo{author}{\bibfnamefont{V.~G.} \bibnamefont{Satorras}},
  \bibinfo{author}{\bibfnamefont{C.-W.} \bibnamefont{Huang}},
  \bibinfo{author}{\bibfnamefont{D.}~\bibnamefont{Zuegner}},
  \bibinfo{author}{\bibfnamefont{M.}~\bibnamefont{Federici}},
  \bibinfo{author}{\bibfnamefont{C.}~\bibnamefont{Clementi}},
  \bibinfo{author}{\bibfnamefont{F.}~\bibnamefont{No{\'e}}},
  \bibinfo{author}{\bibfnamefont{R.}~\bibnamefont{Pinsler}}, \bibnamefont{and}
  \bibinfo{author}{\bibfnamefont{R.}~\bibnamefont{van~den Berg}},
  \emph{\bibinfo{title}{Two for {{One}}: {{Diffusion Models}} and {{Force
  Fields}} for {{Coarse-Grained Molecular Dynamics}}}} (\bibinfo{year}{2023}),
  \eprint{arXiv:2302.00600}.

\bibitem[{\citenamefont{Shui et~al.}(2022)\citenamefont{Shui, Karls, Wen,
  Nikiforov, Tadmor, and Karypis}}]{Shui-22-10}
\bibinfo{author}{\bibfnamefont{Z.}~\bibnamefont{Shui}},
  \bibinfo{author}{\bibfnamefont{D.~S.} \bibnamefont{Karls}},
  \bibinfo{author}{\bibfnamefont{M.}~\bibnamefont{Wen}},
  \bibinfo{author}{\bibfnamefont{I.~A.} \bibnamefont{Nikiforov}},
  \bibinfo{author}{\bibfnamefont{E.~B.} \bibnamefont{Tadmor}},
  \bibnamefont{and} \bibinfo{author}{\bibfnamefont{G.}~\bibnamefont{Karypis}},
  \emph{\bibinfo{title}{Injecting {{Domain Knowledge}} from {{Empirical
  Interatomic Potentials}} to {{Neural Networks}} for {{Predicting Material
  Properties}}}} (\bibinfo{year}{2022}), \eprint{arXiv:2210.08047}.

\bibitem[{\citenamefont{Zhuang et~al.}(2019)\citenamefont{Zhuang, Qi, Duan, Xi,
  Zhu, Zhu, Xiong, and He}}]{Zhuang-19-11}
\bibinfo{author}{\bibfnamefont{F.}~\bibnamefont{Zhuang}},
  \bibinfo{author}{\bibfnamefont{Z.}~\bibnamefont{Qi}},
  \bibinfo{author}{\bibfnamefont{K.}~\bibnamefont{Duan}},
  \bibinfo{author}{\bibfnamefont{D.}~\bibnamefont{Xi}},
  \bibinfo{author}{\bibfnamefont{Y.}~\bibnamefont{Zhu}},
  \bibinfo{author}{\bibfnamefont{H.}~\bibnamefont{Zhu}},
  \bibinfo{author}{\bibfnamefont{H.}~\bibnamefont{Xiong}}, \bibnamefont{and}
  \bibinfo{author}{\bibfnamefont{Q.}~\bibnamefont{He}}, \emph{\bibinfo{title}{A
  {{Comprehensive Survey}} on {{Transfer Learning}}}} (\bibinfo{year}{2019}),
  \eprint{arXiv:1911.02685}.

\bibitem[{\citenamefont{Saenko et~al.}(2010)\citenamefont{Saenko, Kulis, Fritz,
  and Darrell}}]{Saenko-10}
\bibinfo{author}{\bibfnamefont{K.}~\bibnamefont{Saenko}},
  \bibinfo{author}{\bibfnamefont{B.}~\bibnamefont{Kulis}},
  \bibinfo{author}{\bibfnamefont{M.}~\bibnamefont{Fritz}}, \bibnamefont{and}
  \bibinfo{author}{\bibfnamefont{T.}~\bibnamefont{Darrell}}, in
  \emph{\bibinfo{booktitle}{Computer {{Vision}} \textendash{} {{ECCV}} 2010}},
  edited by \bibinfo{editor}{\bibfnamefont{K.}~\bibnamefont{Daniilidis}},
  \bibinfo{editor}{\bibfnamefont{P.}~\bibnamefont{Maragos}}, \bibnamefont{and}
  \bibinfo{editor}{\bibfnamefont{N.}~\bibnamefont{Paragios}}
  (\bibinfo{address}{{Berlin, Heidelberg}}, \bibinfo{year}{2010}), Lecture
  {{Notes}} in {{Computer Science}}, pp. \bibinfo{pages}{213--226}.

\bibitem[{\citenamefont{Vinyals et~al.}(2015)\citenamefont{Vinyals, Toshev,
  Bengio, and Erhan}}]{Vinyals-15-06}
\bibinfo{author}{\bibfnamefont{O.}~\bibnamefont{Vinyals}},
  \bibinfo{author}{\bibfnamefont{A.}~\bibnamefont{Toshev}},
  \bibinfo{author}{\bibfnamefont{S.}~\bibnamefont{Bengio}}, \bibnamefont{and}
  \bibinfo{author}{\bibfnamefont{D.}~\bibnamefont{Erhan}}, in
  \emph{\bibinfo{booktitle}{2015 {IEEE} {Conference} on {Computer} {Vision} and
  {Pattern} {Recognition} ({CVPR})}} (\bibinfo{publisher}{IEEE},
  \bibinfo{address}{Boston, MA, USA}, \bibinfo{year}{2015}), pp.
  \bibinfo{pages}{3156--3164}.

\bibitem[{\citenamefont{Sharma et~al.}(2007)\citenamefont{Sharma, Holmes,
  Santamaria, Irani, Isbell, and Ram}}]{Sharma-07-01}
\bibinfo{author}{\bibfnamefont{M.}~\bibnamefont{Sharma}},
  \bibinfo{author}{\bibfnamefont{M.}~\bibnamefont{Holmes}},
  \bibinfo{author}{\bibfnamefont{J.}~\bibnamefont{Santamaria}},
  \bibinfo{author}{\bibfnamefont{A.}~\bibnamefont{Irani}},
  \bibinfo{author}{\bibfnamefont{C.}~\bibnamefont{Isbell}}, \bibnamefont{and}
  \bibinfo{author}{\bibfnamefont{A.}~\bibnamefont{Ram}},
  \bibinfo{journal}{International Joint Conference on Artificial Intelligence}
  (\bibinfo{year}{2007}).

\bibitem[{\citenamefont{Tang et~al.}(2016)\citenamefont{Tang, Lou, Kleinberg,
  and Wu}}]{Tang-16-04}
\bibinfo{author}{\bibfnamefont{J.}~\bibnamefont{Tang}},
  \bibinfo{author}{\bibfnamefont{T.}~\bibnamefont{Lou}},
  \bibinfo{author}{\bibfnamefont{J.}~\bibnamefont{Kleinberg}},
  \bibnamefont{and} \bibinfo{author}{\bibfnamefont{S.}~\bibnamefont{Wu}},
  \bibinfo{journal}{ACM Transactions on Information Systems}
  \textbf{\bibinfo{volume}{34}}, \bibinfo{pages}{7:1} (\bibinfo{year}{2016}).

\bibitem[{\citenamefont{Smith et~al.}(2019)\citenamefont{Smith, Nebgen,
  Zubatyuk, Lubbers, Devereux, Barros, Tretiak, Isayev, and
  Roitberg}}]{Smith-19-07}
\bibinfo{author}{\bibfnamefont{J.~S.} \bibnamefont{Smith}},
  \bibinfo{author}{\bibfnamefont{B.~T.} \bibnamefont{Nebgen}},
  \bibinfo{author}{\bibfnamefont{R.}~\bibnamefont{Zubatyuk}},
  \bibinfo{author}{\bibfnamefont{N.}~\bibnamefont{Lubbers}},
  \bibinfo{author}{\bibfnamefont{C.}~\bibnamefont{Devereux}},
  \bibinfo{author}{\bibfnamefont{K.}~\bibnamefont{Barros}},
  \bibinfo{author}{\bibfnamefont{S.}~\bibnamefont{Tretiak}},
  \bibinfo{author}{\bibfnamefont{O.}~\bibnamefont{Isayev}}, \bibnamefont{and}
  \bibinfo{author}{\bibfnamefont{A.~E.} \bibnamefont{Roitberg}},
  \bibinfo{journal}{Nat. Commun.} \textbf{\bibinfo{volume}{10}},
  \bibinfo{pages}{2903} (\bibinfo{year}{2019}).

\bibitem[{\citenamefont{Zhang et~al.}(2022{\natexlab{b}})\citenamefont{Zhang,
  Bi, Dai, Jiang, Zhang, and Wang}}]{Zhang-22-09}
\bibinfo{author}{\bibfnamefont{D.}~\bibnamefont{Zhang}},
  \bibinfo{author}{\bibfnamefont{H.}~\bibnamefont{Bi}},
  \bibinfo{author}{\bibfnamefont{F.-Z.} \bibnamefont{Dai}},
  \bibinfo{author}{\bibfnamefont{W.}~\bibnamefont{Jiang}},
  \bibinfo{author}{\bibfnamefont{L.}~\bibnamefont{Zhang}}, \bibnamefont{and}
  \bibinfo{author}{\bibfnamefont{H.}~\bibnamefont{Wang}},
  \emph{\bibinfo{title}{{DPA}-1: {Pretraining} of {Attention}-based {Deep}
  {Potential} {Model} for {Molecular} {Simulation}}}
  (\bibinfo{year}{2022}{\natexlab{b}}), \eprint{arXiv:2208.08236}.

\bibitem[{\citenamefont{Chen et~al.}(2023)\citenamefont{Chen, Lee, Ye,
  Berkelbach, Reichman, and Markland}}]{Chen-23-02}
\bibinfo{author}{\bibfnamefont{M.~S.} \bibnamefont{Chen}},
  \bibinfo{author}{\bibfnamefont{J.}~\bibnamefont{Lee}},
  \bibinfo{author}{\bibfnamefont{H.-Z.} \bibnamefont{Ye}},
  \bibinfo{author}{\bibfnamefont{T.~C.} \bibnamefont{Berkelbach}},
  \bibinfo{author}{\bibfnamefont{D.~R.} \bibnamefont{Reichman}},
  \bibnamefont{and} \bibinfo{author}{\bibfnamefont{T.~E.}
  \bibnamefont{Markland}}, \bibinfo{journal}{J. Chem. Theory Comput.}
  (\bibinfo{year}{2023}).

\bibitem[{\citenamefont{Li et~al.}(2022)\citenamefont{Li, Zhao, Li, Wan, Zhao,
  and Zeng}}]{Li-22-10}
\bibinfo{author}{\bibfnamefont{H.}~\bibnamefont{Li}},
  \bibinfo{author}{\bibfnamefont{X.}~\bibnamefont{Zhao}},
  \bibinfo{author}{\bibfnamefont{S.}~\bibnamefont{Li}},
  \bibinfo{author}{\bibfnamefont{F.}~\bibnamefont{Wan}},
  \bibinfo{author}{\bibfnamefont{D.}~\bibnamefont{Zhao}}, \bibnamefont{and}
  \bibinfo{author}{\bibfnamefont{J.}~\bibnamefont{Zeng}},
  \bibinfo{journal}{iScience} \textbf{\bibinfo{volume}{25}},
  \bibinfo{pages}{105231} (\bibinfo{year}{2022}).

\bibitem[{\citenamefont{Faber et~al.}(2018)\citenamefont{Faber, Christensen,
  Huang, and von Lilienfeld}}]{Faber-18-03}
\bibinfo{author}{\bibfnamefont{F.~A.} \bibnamefont{Faber}},
  \bibinfo{author}{\bibfnamefont{A.~S.} \bibnamefont{Christensen}},
  \bibinfo{author}{\bibfnamefont{B.}~\bibnamefont{Huang}}, \bibnamefont{and}
  \bibinfo{author}{\bibfnamefont{O.~A.} \bibnamefont{von Lilienfeld}},
  \bibinfo{journal}{J. Chem. Phys.} \textbf{\bibinfo{volume}{148}},
  \bibinfo{pages}{241717} (\bibinfo{year}{2018}).

\bibitem[{\citenamefont{Fias et~al.}(2019)\citenamefont{Fias, Chang, and von
  Lilienfeld}}]{Fias-19-01}
\bibinfo{author}{\bibfnamefont{S.}~\bibnamefont{Fias}},
  \bibinfo{author}{\bibfnamefont{K.~Y.~S.} \bibnamefont{Chang}},
  \bibnamefont{and} \bibinfo{author}{\bibfnamefont{O.~A.} \bibnamefont{von
  Lilienfeld}}, \bibinfo{journal}{J. Phys. Chem. Lett.}
  \textbf{\bibinfo{volume}{10}}, \bibinfo{pages}{30} (\bibinfo{year}{2019}).

\bibitem[{\citenamefont{Hornik}(1991)}]{Hornik-91-01}
\bibinfo{author}{\bibfnamefont{K.}~\bibnamefont{Hornik}},
  \bibinfo{journal}{Neural Networks} \textbf{\bibinfo{volume}{4}},
  \bibinfo{pages}{251} (\bibinfo{year}{1991}).

\bibitem[{\citenamefont{Behler and Parrinello}(2007)}]{Behler-07-04}
\bibinfo{author}{\bibfnamefont{J.}~\bibnamefont{Behler}} \bibnamefont{and}
  \bibinfo{author}{\bibfnamefont{M.}~\bibnamefont{Parrinello}},
  \bibinfo{journal}{Phys. Rev. Lett.} \textbf{\bibinfo{volume}{98}},
  \bibinfo{pages}{146401} (\bibinfo{year}{2007}).

\bibitem[{\citenamefont{Bart{\'o}k et~al.}(2013)\citenamefont{Bart{\'o}k,
  Kondor, and Cs{\'a}nyi}}]{Bartok-13-05}
\bibinfo{author}{\bibfnamefont{A.~P.} \bibnamefont{Bart{\'o}k}},
  \bibinfo{author}{\bibfnamefont{R.}~\bibnamefont{Kondor}}, \bibnamefont{and}
  \bibinfo{author}{\bibfnamefont{G.}~\bibnamefont{Cs{\'a}nyi}},
  \bibinfo{journal}{Phys. Rev. B} \textbf{\bibinfo{volume}{87}},
  \bibinfo{pages}{184115} (\bibinfo{year}{2013}).

\bibitem[{\citenamefont{Wang et~al.}(2018)\citenamefont{Wang, Zhang, Han, and
  E}}]{Wang-18-07}
\bibinfo{author}{\bibfnamefont{H.}~\bibnamefont{Wang}},
  \bibinfo{author}{\bibfnamefont{L.}~\bibnamefont{Zhang}},
  \bibinfo{author}{\bibfnamefont{J.}~\bibnamefont{Han}}, \bibnamefont{and}
  \bibinfo{author}{\bibfnamefont{W.}~\bibnamefont{E}},
  \bibinfo{journal}{Comput. Phys. Commun.} \textbf{\bibinfo{volume}{228}},
  \bibinfo{pages}{178} (\bibinfo{year}{2018}).

\bibitem[{\citenamefont{Drautz}(2019)}]{Drautz-19-01}
\bibinfo{author}{\bibfnamefont{R.}~\bibnamefont{Drautz}},
  \bibinfo{journal}{Phys. Rev. B} \textbf{\bibinfo{volume}{99}},
  \bibinfo{pages}{014104} (\bibinfo{year}{2019}).

\bibitem[{\citenamefont{Sch\"u{}tt et~al.}(2018)\citenamefont{Sch\"u{}tt,
  Sauceda, Kindermans, Tkatchenko, and M\"u{}ller}}]{Schuett-18-03}
\bibinfo{author}{\bibfnamefont{K.~T.} \bibnamefont{Sch\"u{}tt}},
  \bibinfo{author}{\bibfnamefont{H.~E.} \bibnamefont{Sauceda}},
  \bibinfo{author}{\bibfnamefont{P.-J.} \bibnamefont{Kindermans}},
  \bibinfo{author}{\bibfnamefont{A.}~\bibnamefont{Tkatchenko}},
  \bibnamefont{and} \bibinfo{author}{\bibfnamefont{K.-R.}
  \bibnamefont{M\"u{}ller}}, \bibinfo{journal}{J. Chem. Phys.}
  \textbf{\bibinfo{volume}{148}}, \bibinfo{pages}{241722}
  (\bibinfo{year}{2018}).

\bibitem[{\citenamefont{Unke and Meuwly}(2019)}]{Unke-19-05}
\bibinfo{author}{\bibfnamefont{O.~T.} \bibnamefont{Unke}} \bibnamefont{and}
  \bibinfo{author}{\bibfnamefont{M.}~\bibnamefont{Meuwly}},
  \bibinfo{journal}{J. Chem. Theory Comput.} \textbf{\bibinfo{volume}{15}},
  \bibinfo{pages}{3678} (\bibinfo{year}{2019}).

\bibitem[{\citenamefont{Ko et~al.}(2021)\citenamefont{Ko, Finkler, Goedecker,
  and Behler}}]{Ko-21-01}
\bibinfo{author}{\bibfnamefont{T.~W.} \bibnamefont{Ko}},
  \bibinfo{author}{\bibfnamefont{J.~A.} \bibnamefont{Finkler}},
  \bibinfo{author}{\bibfnamefont{S.}~\bibnamefont{Goedecker}},
  \bibnamefont{and} \bibinfo{author}{\bibfnamefont{J.}~\bibnamefont{Behler}},
  \bibinfo{journal}{Nat. Commun.} \textbf{\bibinfo{volume}{12}},
  \bibinfo{pages}{398} (\bibinfo{year}{2021}).

\bibitem[{\citenamefont{Sch{\"u}tt et~al.}(2021)\citenamefont{Sch{\"u}tt, Unke,
  and Gastegger}}]{Schuett-21-06}
\bibinfo{author}{\bibfnamefont{K.~T.} \bibnamefont{Sch{\"u}tt}},
  \bibinfo{author}{\bibfnamefont{O.~T.} \bibnamefont{Unke}}, \bibnamefont{and}
  \bibinfo{author}{\bibfnamefont{M.}~\bibnamefont{Gastegger}},
  \emph{\bibinfo{title}{Equivariant message passing for the prediction of
  tensorial properties and molecular spectra}} (\bibinfo{year}{2021}),
  \eprint{arXiv:2102.03150}.

\bibitem[{\citenamefont{Batzner et~al.}(2022)\citenamefont{Batzner, Musaelian,
  Sun, Geiger, Mailoa, Kornbluth, Molinari, Smidt, and
  Kozinsky}}]{Batzner-22-05}
\bibinfo{author}{\bibfnamefont{S.}~\bibnamefont{Batzner}},
  \bibinfo{author}{\bibfnamefont{A.}~\bibnamefont{Musaelian}},
  \bibinfo{author}{\bibfnamefont{L.}~\bibnamefont{Sun}},
  \bibinfo{author}{\bibfnamefont{M.}~\bibnamefont{Geiger}},
  \bibinfo{author}{\bibfnamefont{J.~P.} \bibnamefont{Mailoa}},
  \bibinfo{author}{\bibfnamefont{M.}~\bibnamefont{Kornbluth}},
  \bibinfo{author}{\bibfnamefont{N.}~\bibnamefont{Molinari}},
  \bibinfo{author}{\bibfnamefont{T.~E.} \bibnamefont{Smidt}}, \bibnamefont{and}
  \bibinfo{author}{\bibfnamefont{B.}~\bibnamefont{Kozinsky}},
  \bibinfo{journal}{Nat. Commun.} \textbf{\bibinfo{volume}{13}},
  \bibinfo{pages}{2453} (\bibinfo{year}{2022}).

\bibitem[{\citenamefont{Batatia
  et~al.}(2022{\natexlab{a}})\citenamefont{Batatia, Kovacs, Simm, Ortner, and
  Csanyi}}]{Batatia-22-10}
\bibinfo{author}{\bibfnamefont{I.}~\bibnamefont{Batatia}},
  \bibinfo{author}{\bibfnamefont{D.~P.} \bibnamefont{Kovacs}},
  \bibinfo{author}{\bibfnamefont{G.~N.~C.} \bibnamefont{Simm}},
  \bibinfo{author}{\bibfnamefont{C.}~\bibnamefont{Ortner}}, \bibnamefont{and}
  \bibinfo{author}{\bibfnamefont{G.}~\bibnamefont{Csanyi}}, in
  \emph{\bibinfo{booktitle}{Advances in Neural Information Processing
  Systems}}, edited by \bibinfo{editor}{\bibfnamefont{A.~H.} \bibnamefont{Oh}},
  \bibinfo{editor}{\bibfnamefont{A.}~\bibnamefont{Agarwal}},
  \bibinfo{editor}{\bibfnamefont{D.}~\bibnamefont{Belgrave}}, \bibnamefont{and}
  \bibinfo{editor}{\bibfnamefont{K.}~\bibnamefont{Cho}}
  (\bibinfo{year}{2022}{\natexlab{a}}), vol.~\bibinfo{volume}{35}, pp.
  \bibinfo{pages}{11423--11436}.

\bibitem[{\citenamefont{Batatia
  et~al.}(2022{\natexlab{b}})\citenamefont{Batatia, Batzner, Kov\'a{}cs,
  Musaelian, Simm, Drautz, Ortner, Kozinsky, and Cs\'a{}nyi}}]{Batatia-22-05}
\bibinfo{author}{\bibfnamefont{I.}~\bibnamefont{Batatia}},
  \bibinfo{author}{\bibfnamefont{S.}~\bibnamefont{Batzner}},
  \bibinfo{author}{\bibfnamefont{D.~P.} \bibnamefont{Kov\'a{}cs}},
  \bibinfo{author}{\bibfnamefont{A.}~\bibnamefont{Musaelian}},
  \bibinfo{author}{\bibfnamefont{G.~N.~C.} \bibnamefont{Simm}},
  \bibinfo{author}{\bibfnamefont{R.}~\bibnamefont{Drautz}},
  \bibinfo{author}{\bibfnamefont{C.}~\bibnamefont{Ortner}},
  \bibinfo{author}{\bibfnamefont{B.}~\bibnamefont{Kozinsky}}, \bibnamefont{and}
  \bibinfo{author}{\bibfnamefont{G.}~\bibnamefont{Cs\'a{}nyi}},
  \emph{\bibinfo{title}{{The Design Space of E(3)-Equivariant Atom-Centered
  Interatomic Potentials}}} (\bibinfo{year}{2022}{\natexlab{b}}),
  \eprint{arXiv:2205.06643}.

\bibitem[{\citenamefont{Simeon and {de Fabritiis}}(2023)}]{Simeon-23-06}
\bibinfo{author}{\bibfnamefont{G.}~\bibnamefont{Simeon}} \bibnamefont{and}
  \bibinfo{author}{\bibfnamefont{G.}~\bibnamefont{{de Fabritiis}}},
  \emph{\bibinfo{title}{{{TensorNet}}: {{Cartesian Tensor Representations}} for
  {{Efficient Learning}} of {{Molecular Potentials}}}} (\bibinfo{year}{2023}),
  \eprint{arXiv:2306.06482}.

\bibitem[{\citenamefont{Liu et~al.}(2021)\citenamefont{Liu, Agarwal, and
  Venkataraman}}]{Liu-21-04}
\bibinfo{author}{\bibfnamefont{Y.}~\bibnamefont{Liu}},
  \bibinfo{author}{\bibfnamefont{S.}~\bibnamefont{Agarwal}}, \bibnamefont{and}
  \bibinfo{author}{\bibfnamefont{S.}~\bibnamefont{Venkataraman}},
  \emph{\bibinfo{title}{{{AutoFreeze}}: {{Automatically Freezing Model Blocks}}
  to {{Accelerate Fine-tuning}}}} (\bibinfo{year}{2021}),
  \eprint{arXiv:2102.01386}.

\bibitem[{\citenamefont{Frankle and Carbin}(2019)}]{Frankle-19-03}
\bibinfo{author}{\bibfnamefont{J.}~\bibnamefont{Frankle}} \bibnamefont{and}
  \bibinfo{author}{\bibfnamefont{M.}~\bibnamefont{Carbin}},
  \emph{\bibinfo{title}{The {{Lottery Ticket Hypothesis}}: {{Finding Sparse}},
  {{Trainable Neural Networks}}}} (\bibinfo{year}{2019}),
  \eprint{arXiv:1803.03635}.

\bibitem[{\citenamefont{Howard and Ruder}(2018)}]{Howard-18-05}
\bibinfo{author}{\bibfnamefont{J.}~\bibnamefont{Howard}} \bibnamefont{and}
  \bibinfo{author}{\bibfnamefont{S.}~\bibnamefont{Ruder}},
  \emph{\bibinfo{title}{Universal {{Language Model Fine-tuning}} for {{Text
  Classification}}}} (\bibinfo{year}{2018}), \eprint{arXiv:1801.06146}.

\bibitem[{\citenamefont{Thompson et~al.}(2022)\citenamefont{Thompson, Aktulga,
  Berger, Bolintineanu, Brown, Crozier, {in 't Veld}, Kohlmeyer, Moore, Nguyen
  et~al.}}]{Thompson-22-02}
\bibinfo{author}{\bibfnamefont{A.~P.} \bibnamefont{Thompson}},
  \bibinfo{author}{\bibfnamefont{H.~M.} \bibnamefont{Aktulga}},
  \bibinfo{author}{\bibfnamefont{R.}~\bibnamefont{Berger}},
  \bibinfo{author}{\bibfnamefont{D.~S.} \bibnamefont{Bolintineanu}},
  \bibinfo{author}{\bibfnamefont{W.~M.} \bibnamefont{Brown}},
  \bibinfo{author}{\bibfnamefont{P.~S.} \bibnamefont{Crozier}},
  \bibinfo{author}{\bibfnamefont{P.~J.} \bibnamefont{{in 't Veld}}},
  \bibinfo{author}{\bibfnamefont{A.}~\bibnamefont{Kohlmeyer}},
  \bibinfo{author}{\bibfnamefont{S.~G.} \bibnamefont{Moore}},
  \bibinfo{author}{\bibfnamefont{T.~D.} \bibnamefont{Nguyen}},
  \bibnamefont{et~al.}, \bibinfo{journal}{Comput. Phys. Commun.}
  \textbf{\bibinfo{volume}{271}}, \bibinfo{pages}{108171}
  (\bibinfo{year}{2022}).

\bibitem[{\citenamefont{Deringer and Cs{\'a}nyi}(2017)}]{Deringer-17-03}
\bibinfo{author}{\bibfnamefont{V.~L.} \bibnamefont{Deringer}} \bibnamefont{and}
  \bibinfo{author}{\bibfnamefont{G.}~\bibnamefont{Cs{\'a}nyi}},
  \bibinfo{journal}{Phys. Rev. B} \textbf{\bibinfo{volume}{95}},
  \bibinfo{pages}{094203} (\bibinfo{year}{2017}).

\bibitem[{\citenamefont{Bart{\'o}k et~al.}(2010)\citenamefont{Bart{\'o}k,
  Payne, Kondor, and Cs{\'a}nyi}}]{Bartok-10-04}
\bibinfo{author}{\bibfnamefont{A.~P.} \bibnamefont{Bart{\'o}k}},
  \bibinfo{author}{\bibfnamefont{M.~C.} \bibnamefont{Payne}},
  \bibinfo{author}{\bibfnamefont{R.}~\bibnamefont{Kondor}}, \bibnamefont{and}
  \bibinfo{author}{\bibfnamefont{G.}~\bibnamefont{Cs{\'a}nyi}},
  \bibinfo{journal}{Phys. Rev. Lett.} \textbf{\bibinfo{volume}{104}},
  \bibinfo{pages}{136403} (\bibinfo{year}{2010}).

\bibitem[{\citenamefont{Rowe et~al.}(2020)\citenamefont{Rowe, Deringer,
  Gasparotto, Cs{\'a}nyi, and Michaelides}}]{Rowe-20-07}
\bibinfo{author}{\bibfnamefont{P.}~\bibnamefont{Rowe}},
  \bibinfo{author}{\bibfnamefont{V.~L.} \bibnamefont{Deringer}},
  \bibinfo{author}{\bibfnamefont{P.}~\bibnamefont{Gasparotto}},
  \bibinfo{author}{\bibfnamefont{G.}~\bibnamefont{Cs{\'a}nyi}},
  \bibnamefont{and}
  \bibinfo{author}{\bibfnamefont{A.}~\bibnamefont{Michaelides}},
  \bibinfo{journal}{J. Chem. Phys.} \textbf{\bibinfo{volume}{153}},
  \bibinfo{pages}{034702} (\bibinfo{year}{2020}).

\bibitem[{\citenamefont{Qamar et~al.}(2023)\citenamefont{Qamar, Mrovec,
  Lysogorskiy, Bochkarev, and Drautz}}]{Qamar-23-06}
\bibinfo{author}{\bibfnamefont{M.}~\bibnamefont{Qamar}},
  \bibinfo{author}{\bibfnamefont{M.}~\bibnamefont{Mrovec}},
  \bibinfo{author}{\bibfnamefont{Y.}~\bibnamefont{Lysogorskiy}},
  \bibinfo{author}{\bibfnamefont{A.}~\bibnamefont{Bochkarev}},
  \bibnamefont{and} \bibinfo{author}{\bibfnamefont{R.}~\bibnamefont{Drautz}},
  \bibinfo{journal}{J. Chem. Theory Comput.}, DOI:
  10.1021/acs.jctc.2c01149 (\bibinfo{year}{2023}).

\bibitem[{\citenamefont{Los and Fasolino}(2003)}]{Los-03-07}
\bibinfo{author}{\bibfnamefont{J.~H.} \bibnamefont{Los}} \bibnamefont{and}
  \bibinfo{author}{\bibfnamefont{A.}~\bibnamefont{Fasolino}},
  \bibinfo{journal}{Phys. Rev. B} \textbf{\bibinfo{volume}{68}},
  \bibinfo{pages}{024107} (\bibinfo{year}{2003}).

\bibitem[{\citenamefont{Bazant and Kaxiras}(1996)}]{Bazant-96-11}
\bibinfo{author}{\bibfnamefont{M.~Z.} \bibnamefont{Bazant}} \bibnamefont{and}
  \bibinfo{author}{\bibfnamefont{E.}~\bibnamefont{Kaxiras}},
  \bibinfo{journal}{Phys. Rev. Lett.} \textbf{\bibinfo{volume}{77}},
  \bibinfo{pages}{4370} (\bibinfo{year}{1996}).

\bibitem[{\citenamefont{Marks}(2000)}]{Marks-00-12}
\bibinfo{author}{\bibfnamefont{N.~A.} \bibnamefont{Marks}},
  \bibinfo{journal}{Phys. Rev. B} \textbf{\bibinfo{volume}{63}},
  \bibinfo{pages}{035401} (\bibinfo{year}{2000}).

\bibitem[{\citenamefont{Bart\'o{}k et~al.}(2018)\citenamefont{Bart\'o{}k,
  Kermode, Bernstein, and Cs\'a{}nyi}}]{Bartok-18-12}
\bibinfo{author}{\bibfnamefont{A.~P.} \bibnamefont{Bart\'o{}k}},
  \bibinfo{author}{\bibfnamefont{J.}~\bibnamefont{Kermode}},
  \bibinfo{author}{\bibfnamefont{N.}~\bibnamefont{Bernstein}},
  \bibnamefont{and}
  \bibinfo{author}{\bibfnamefont{G.}~\bibnamefont{Cs\'a{}nyi}},
  \bibinfo{journal}{Phys. Rev. X} \textbf{\bibinfo{volume}{8}},
  \bibinfo{pages}{041048} (\bibinfo{year}{2018}).

\bibitem[{\citenamefont{Kingma and Ba}(2017)}]{Kingma-17-01}
\bibinfo{author}{\bibfnamefont{D.~P.} \bibnamefont{Kingma}} \bibnamefont{and}
  \bibinfo{author}{\bibfnamefont{J.}~\bibnamefont{Ba}},
  \emph{\bibinfo{title}{Adam: {{A Method}} for {{Stochastic Optimization}}}}
  (\bibinfo{year}{2017}), \eprint{arXiv:1412.6980}.

\bibitem[{\citenamefont{Karls}(2019)}]{LCBOP-KIM-19}
\bibinfo{author}{\bibfnamefont{D.~S.} \bibnamefont{Karls}},
  \emph{\bibinfo{title}{{LAMMPS} {LCBOP} potential for {C} developed by {L}os
  and {F}asolino (2003) v000}}, \bibinfo{howpublished}{OpenKIM}
  (\bibinfo{year}{2019}).

\bibitem[{\citenamefont{Karls et~al.}(2018)\citenamefont{Karls, Justo, Bazant,
  Kaxiras, Bulatov, and Yip}}]{EDIP-Kim-18}
\bibinfo{author}{\bibfnamefont{D.~S.} \bibnamefont{Karls}},
  \bibinfo{author}{\bibfnamefont{J.~F.} \bibnamefont{Justo}},
  \bibinfo{author}{\bibfnamefont{M.~Z.} \bibnamefont{Bazant}},
  \bibinfo{author}{\bibfnamefont{E.}~\bibnamefont{Kaxiras}},
  \bibinfo{author}{\bibfnamefont{V.~V.} \bibnamefont{Bulatov}},
  \bibnamefont{and} \bibinfo{author}{\bibfnamefont{S.}~\bibnamefont{Yip}},
  \emph{\bibinfo{title}{{E}nvironment-{D}ependent {I}nteratomic {P}otential
  ({EDIP}) model driver v002}}, \bibinfo{howpublished}{OpenKIM}
  (\bibinfo{year}{2018}).

\bibitem[{\citenamefont{Tadmor et~al.}(2011)\citenamefont{Tadmor, Elliott,
  Sethna, Miller, and Becker}}]{Tadmor-11}
\bibinfo{author}{\bibfnamefont{E.~B.} \bibnamefont{Tadmor}},
  \bibinfo{author}{\bibfnamefont{R.~S.} \bibnamefont{Elliott}},
  \bibinfo{author}{\bibfnamefont{J.~P.} \bibnamefont{Sethna}},
  \bibinfo{author}{\bibfnamefont{R.~E.} \bibnamefont{Miller}},
  \bibnamefont{and} \bibinfo{author}{\bibfnamefont{C.~A.}
  \bibnamefont{Becker}}, \bibinfo{journal}{{JOM}}
  \textbf{\bibinfo{volume}{63}}, \bibinfo{pages}{17} (\bibinfo{year}{2011}).

\bibitem[{\citenamefont{Elliott and Tadmor}(2011)}]{Elliott-11}
\bibinfo{author}{\bibfnamefont{R.~S.} \bibnamefont{Elliott}} \bibnamefont{and}
  \bibinfo{author}{\bibfnamefont{E.~B.} \bibnamefont{Tadmor}},
  \emph{\bibinfo{title}{{K}nowledgebase of {I}nteratomic {M}odels ({KIM})
  application programming interface ({API})}} (\bibinfo{year}{2011}).

\bibitem[{\citenamefont{Bochkarev et~al.}(2022)\citenamefont{Bochkarev,
  Lysogorskiy, Menon, Qamar, Mrovec, and Drautz}}]{Bochkarev-22}
\bibinfo{author}{\bibfnamefont{A.}~\bibnamefont{Bochkarev}},
  \bibinfo{author}{\bibfnamefont{Y.}~\bibnamefont{Lysogorskiy}},
  \bibinfo{author}{\bibfnamefont{S.}~\bibnamefont{Menon}},
  \bibinfo{author}{\bibfnamefont{M.}~\bibnamefont{Qamar}},
  \bibinfo{author}{\bibfnamefont{M.}~\bibnamefont{Mrovec}}, \bibnamefont{and}
  \bibinfo{author}{\bibfnamefont{R.}~\bibnamefont{Drautz}},
  \bibinfo{journal}{Phys. Rev. Mater.} \textbf{\bibinfo{volume}{6}},
  \bibinfo{pages}{013804} (\bibinfo{year}{2022}).

\bibitem[{\citenamefont{Lysogorskiy et~al.}(2021)\citenamefont{Lysogorskiy,
  Oord, Bochkarev, Menon, Rinaldi, Hammerschmidt, Mrovec, Thompson, Cs{\'a}nyi,
  Ortner et~al.}}]{Lysogorskiy-21}
\bibinfo{author}{\bibfnamefont{Y.}~\bibnamefont{Lysogorskiy}},
  \bibinfo{author}{\bibfnamefont{C.~v.~d.} \bibnamefont{Oord}},
  \bibinfo{author}{\bibfnamefont{A.}~\bibnamefont{Bochkarev}},
  \bibinfo{author}{\bibfnamefont{S.}~\bibnamefont{Menon}},
  \bibinfo{author}{\bibfnamefont{M.}~\bibnamefont{Rinaldi}},
  \bibinfo{author}{\bibfnamefont{T.}~\bibnamefont{Hammerschmidt}},
  \bibinfo{author}{\bibfnamefont{M.}~\bibnamefont{Mrovec}},
  \bibinfo{author}{\bibfnamefont{A.}~\bibnamefont{Thompson}},
  \bibinfo{author}{\bibfnamefont{G.}~\bibnamefont{Cs{\'a}nyi}},
  \bibinfo{author}{\bibfnamefont{C.}~\bibnamefont{Ortner}},
  \bibnamefont{et~al.}, \bibinfo{journal}{npj Comput. Mater.}
  \textbf{\bibinfo{volume}{7}}, \bibinfo{pages}{97} (\bibinfo{year}{2021}).

\bibitem[{\citenamefont{Cs{\'a}nyi et~al.}(2007)\citenamefont{Cs{\'a}nyi,
  Winfield, Kermode, De~Vita, Comisso, Bernstein, and Payne}}]{Csanyi-07}
\bibinfo{author}{\bibfnamefont{G.}~\bibnamefont{Cs{\'a}nyi}},
  \bibinfo{author}{\bibfnamefont{S.}~\bibnamefont{Winfield}},
  \bibinfo{author}{\bibfnamefont{J.~R.} \bibnamefont{Kermode}},
  \bibinfo{author}{\bibfnamefont{A.}~\bibnamefont{De~Vita}},
  \bibinfo{author}{\bibfnamefont{A.}~\bibnamefont{Comisso}},
  \bibinfo{author}{\bibfnamefont{N.}~\bibnamefont{Bernstein}},
  \bibnamefont{and} \bibinfo{author}{\bibfnamefont{M.~C.} \bibnamefont{Payne}},
  \bibinfo{journal}{IoP Comput. Phys. Newsletter Spring 2007}
  (\bibinfo{year}{2007}).

\bibitem[{\citenamefont{Kermode}(2020)}]{Kermode-20-03}
\bibinfo{author}{\bibfnamefont{J.~R.} \bibnamefont{Kermode}},
  \bibinfo{journal}{Journal of Physics: Condensed Matter}
  \textbf{\bibinfo{volume}{32}}, \bibinfo{pages}{305901}
  (\bibinfo{year}{2020}).

\bibitem[{\citenamefont{McInnes et~al.}(2020)\citenamefont{McInnes, Healy, and
  Melville}}]{McInnes-20-09}
\bibinfo{author}{\bibfnamefont{L.}~\bibnamefont{McInnes}},
  \bibinfo{author}{\bibfnamefont{J.}~\bibnamefont{Healy}}, \bibnamefont{and}
  \bibinfo{author}{\bibfnamefont{J.}~\bibnamefont{Melville}},
  \emph{\bibinfo{title}{{{UMAP}}: {{Uniform Manifold Approximation}} and
  {{Projection}} for {{Dimension Reduction}}}} (\bibinfo{year}{2020}),
  \eprint{arXiv:1802.03426}.

\bibitem[{\citenamefont{Stukowski}(2009)}]{OVITO}
\bibinfo{author}{\bibfnamefont{A.}~\bibnamefont{Stukowski}},
  \bibinfo{journal}{Model. Simul. Mater. Sci. Eng.}
  \textbf{\bibinfo{volume}{18}}, \bibinfo{pages}{015012}
  (\bibinfo{year}{2009}).

\bibitem[{\citenamefont{Powles et~al.}(2009)\citenamefont{Powles, Marks, and
  Lau}}]{Powles-09-02}
\bibinfo{author}{\bibfnamefont{R.~C.} \bibnamefont{Powles}},
  \bibinfo{author}{\bibfnamefont{N.~A.} \bibnamefont{Marks}}, \bibnamefont{and}
  \bibinfo{author}{\bibfnamefont{D.~W.~M.} \bibnamefont{Lau}},
  \bibinfo{journal}{Phys. Rev. B} \textbf{\bibinfo{volume}{79}},
  \bibinfo{pages}{075430} (\bibinfo{year}{2009}).

\bibitem[{\citenamefont{de~Tomas et~al.}(2016)\citenamefont{de~Tomas,
  Suarez-Martinez, and Marks}}]{DeTomas-16-11}
\bibinfo{author}{\bibfnamefont{C.}~\bibnamefont{de~Tomas}},
  \bibinfo{author}{\bibfnamefont{I.}~\bibnamefont{Suarez-Martinez}},
  \bibnamefont{and} \bibinfo{author}{\bibfnamefont{N.~A.} \bibnamefont{Marks}},
  \bibinfo{journal}{Carbon} \textbf{\bibinfo{volume}{109}},
  \bibinfo{pages}{681} (\bibinfo{year}{2016}).

\end{thebibliography}

\end{document}